\documentstyle[preprint,eqsecnum,epsfig,prd,aps,floats]{revtex}
\begin{document}

\newcommand{\nl}{\nonumber \\}
\newcommand{\eq}[1]{Eq.~(\ref{#1})}
\newcommand{\npol}{\bigtriangleup}
\newcommand{\dpol}{\bigtriangledown}
\newcommand{\cstar}{{\bf C}^\ast}

\preprint{\vbox{\hbox{CALT-68-2162}
                \hbox{hep-th/9803033}  }}
 
\title{A note on brane boxes at finite string coupling}
 
\author{Eric G. Gimon\thanks{email: egimon@theory.caltech.edu} and
	 Martin Gremm\thanks{email: gremm@theory.caltech.edu}}
 
\address{California Institute of Technology, Pasadena, CA 91125}

\maketitle

\begin{abstract}
We consider ${\cal N}=1$ supersymmetric $SU(N_c)$ gauge theories, using 
the type IIB brane construction recently proposed by Hanany and Zaffaroni.
At non-zero string coupling, we find that the bending of branes imposes
consistency
conditions that allow only non-anomalous gauge theories with stable
vacua, i.e., $N_f \ge N_c$, to be constructed. We find qualitative differences
between the brane configurations for $N_f \le 3N_c$ and $N_f > 3N_c$,
corresponding to asymptotically free and infrared free theories respectively.
We also discuss some properties of the brane configurations that may be 
relevant to constructing Seiberg's duality in this framework.
\end{abstract}

\newpage

\section{Introduction}
Many field theories with varying numbers of supersymmetries and in various
dimensions can be realized on the common directions of type II branes. 
For a comprehensive review and references see \cite{review}.
In four-dimensional ${\cal N}=1$ theories the problem arises that not all
global symmetries
are fully realized. Specifically, for an $SU(N_c)$ theory with $N_f$ massless
flavors one expects $SU(N_f)_L \times SU(N_f)_R$ chiral symmetry. However,
in the IIA brane construction of this theory only the diagonal $SU(N_f)$ is
visible. This problem was first addressed in \cite{brodie} where it was 
pointed out that the full chiral symmetry can be restored under certain 
conditions. Subsequently, many theories with full chiral symmetry and chiral
matter content were constructed \cite{ch1,ch2,ch3,ch4,ch5}.

More recently Hanany and Zaffaroni \cite{hz} proposed a construction of
chiral gauge theories using grids of NS fivebranes
as well as D5 branes in type IIB string theory. These constructions have the 
advantage that the full chiral symmetry is manifest. They also provide a method
of constructing genuinely chiral theories, i.e., theories for which no mass term
can be added to the superpotential. In Ref. \cite{hz} the authors
consider these brane constructions at zero string coupling. In this case
all branes meet at right angles or are parallel. At zero coupling there does
not seem to be any obstacle in
constructing anomalous gauge theories, such as an $SU(N_c)$ with only 
anti-fundamentals.

In this note, we take a first step towards an analysis of
these brane configurations at finite string coupling.
Once the string coupling is switched on, there are certain restrictions on the
angles at which NS fivebranes and D5 branes
can meet. This is quite familiar from the study of $(p,q)$-webs of fivebranes
\cite{kol}. Past experience has shown that consistency conditions in string
theory usually enforce such field theory requirements as the vanishing of
anomalies.
In this note we analyze the simplest case discussed in \cite{hz}, an $SU(N_c)$
gauge theory with fundamental flavors, at finite string coupling. 
We propose consistency conditions based on the bending of the branes and show
that they guarantee that the gauge theories are non-anomalous. They also
restrict the number of flavors to $N_f \ge N_c$, which excludes theories with
instable vacua \cite{ads,seiberg}. There are qualitative differences between the
brane constructions of $SU(N_c)$ theories with $N_f \le 3N_c$, which are
asymptotically
free, and the theories with $N_f > 3N_c$, which are free in the infrared.
The consistency conditions also help to explain the origin of some
puzzling properties of the brane configurations at zero string coupling.
These problems appear when we try to find Seiberg's duality \cite{ads,seiberg}
by moving the branes in the same way as in the type IIA construction of these
$SU(N_c)$ theories \cite{kut}.
We show two examples of these difficulties and demonstrate that the consistency
conditions may offer a way to resolve them.

\begin{figure}
\begin{center}
\centerline{\epsfxsize=7.cm  \epsfbox{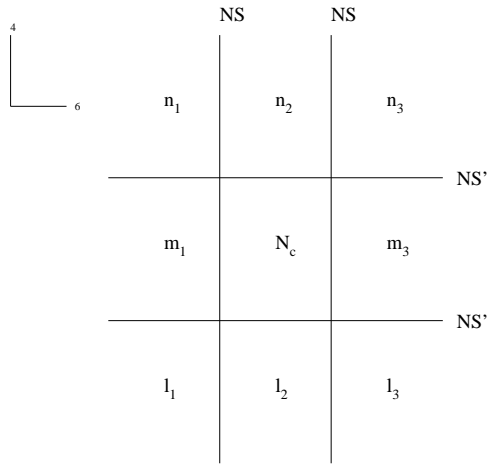}}
\caption{The brane configuration for $SU(N_c)$ with arbitrary numbers of
fundamentals and anti-fundamentals. The indicated number of D5 branes fill the 
the boxes.}
\label{fig1}
\end{center}
\end{figure}

\section{boxes at finite coupling}

Following \cite{hz}, we construct a 3+1 dimensional $SU(N_c)$
gauge theory using type IIB
five-branes with world volumes: NS(012345), NS'(012367), D5(012346).
The NS and NS' branes bound the D5 branes
in the 4 and 6 direction as shown in Fig.~\ref{fig1}.
If we stack $N_c$ D5 branes on top of each other in the finite box we get the
desired
$SU(N_c)$ gauge theory. Matter in the fundamental representation can be 
included by placing additional D5 branes in the semi-infinite boxes surrounding
the box with the $N_c$ finite branes. As indicated in the figure, we place
an arbitrary number of semi-infinite D5 branes in each of these boxes. At
finite string coupling there will be consistency conditions that restrict 
these numbers.  We can determine the (chiral) matter
content of this theory using the rules listed in \cite{hz}.

\begin{figure}
\begin{center}
\centerline{\epsfxsize=15.cm \epsfysize=7.cm \epsfbox{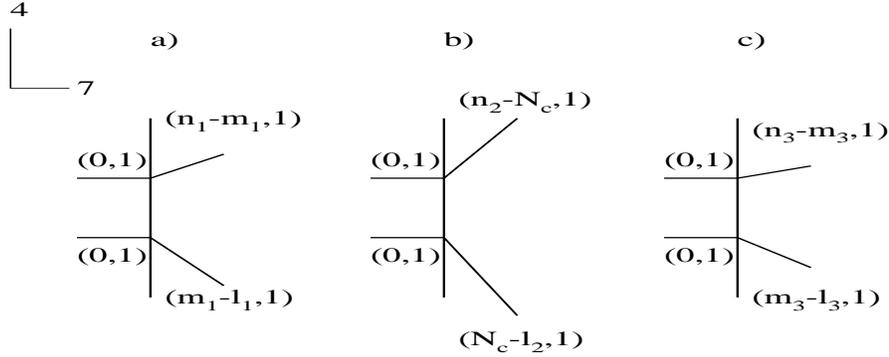}}
\caption{Three slices  showing the $x_4-x_7$ plane: a) to the left of the NS
branes, b) between, c) to the right of the NS branes. The lines labeled $(0,1)$
are the NS' branes and the vertical lines are the stacks of D5 branes.}
\label{fig2}
\end{center}
\end{figure}

The angle at which the D5 and NS branes
intersect is given by
\begin{equation}\label{angle}
\Delta x:\Delta y = p + \tau q,
\end{equation}
where $\tau = i/g_s+\chi/(2\pi)$. (See \cite{kol,hz} for details.)
This condition ensures that no additional
supersymmetries are broken. For $g_s=0$, $\chi=0$
all branes meet at right angles,
but for finite string coupling the angles are determined by the $(p,q)$
charges of the branes involved. The arguments in this section do not depend on
the value of the string coupling as long as it is not zero.
Fig.~\ref{fig2} shows three slices in the $x_4 - x_7$
plane through the configuration in Fig.~\ref{fig1}.
Figures \ref{fig2}a,b,c show $x_6$
positions to the left of the two NS branes, between the NS branes, and to the
right of the NS branes respectively. The angles are determined by the
number of D5 branes ending on each NS' brane from above and below. 
For an arbitrary number of D5 branes in each of the semi-infinite boxes, the
six angles in Fig.~\ref{fig2} will in general be different. 
This requires that far out in the $x_7$ direction the NS' brane is either
discontinuous in the $x_6$ direction or, if the discontinuity is smoothed
out, violates \eq{angle} at least for some $x_6$ positions.
There is an alternative way of looking at this. To the right of the D5 branes
in Fig.~\ref{fig2},
the NS' branes look like $(p,1)$ branes with $p$ depending on the $x_6$
position. There are smooth paths along the NS' branes to the right of the
D5 branes in Fig.~\ref{fig2}, that connect patches with different $p$'s.
Since the $(p,q)$
charges are globally defined quantities, this is not necessarily inconsistent.
However, far away from the finite D5 branes, we expect the angle of the $(p,q)$
brane
to be given by its charges. For arbitrary charges, the $(p,q)$ brane would have
different $p$ charges far out in the $\pm x_6$ direction.
We take these considerations to indicate that the three angles involving
each of the  NS' branes have
to be identical. This imposes the following restrictions on the numbers of
branes we can put in the semi-infinite boxes:
\begin{eqnarray}
n_1 - m_1 = n_2 - N_c = n_3 - m_3,\nl
l_1 - m_1 = l_2 - N_c = l_3 - m_3.
\end{eqnarray}

We can make an identical argument for the bending of the NS branes by
considering slices in the $x_6 - x_5$ plane at different $x_4$ positions.
In order for the NS branes to bend in a consistent way, the numbers of
semi-infinite D5 branes must satisfy
\begin{eqnarray}
n_1-n_2 = m_1-N_c = l_1-l_2,\nl
n_3-n_2 = m_3-N_c = l_3-l_2.
\end{eqnarray}
It is interesting to note that this second set of conditions is linearly
dependent on the first and does not impose any new constraints. 
Solving these consistency conditions gives
\begin{eqnarray}\label{soln}
m_1 &=& n_1-n_2+N_c,\nl
m_3 &=& -n_2+n_3+N_c,\\
l_1 &=& l_3+n_1-n_3,\nl
l_2 &=& l_3+n_2-n_3,\nonumber 
\end{eqnarray}
with $n_1,n_2,n_3,l_3$ arbitrary, and $N_c$ the number of colors. The rules
given in \cite{hz} can be used 
to find the matter content of this theory. The consistency conditions
guarantee that the $SU(N_c)$ theory is anomaly free. There are $l_3+n_1+N_c$
flavors of fundamentals and anti-fundamentals. Note that the minimum
number of flavors in a consistent theory is $N_c$. This agrees with the 
field theory analysis which shows that there is no stable vacuum for $N_f<N_c$
\cite{ads,seiberg}. There is also a a qualitative change in the brane
construction as the number of flavors exceeds $3N_c$. To
see this, consider a theory with $m_1=m_3=N_c$ and $n_1=n_2=n_3=N_f$
and the other semi-infinite boxes empty. This 
configuration satisfies our consistency conditions. It gives rise to an
$SU(N_c)$ gauge theory with $N_f+N_c$ flavors. For $N_f<N_c$ the $(p,q)$ branes
bend away from the stacks of $N_c$ D5 branes as shown in Fig.~\ref{fig2}, but
for $N_f>N_c$ both $(p,q)$ branes bend in the same directions in the $x_4-x_7$
plane. If $N_f>2N_c$, the two $(p,q)$ branes meet somewhere to the right of
the D5 branes in Fig.~\ref{fig2}. There are two inequivalent choices
for the brane configurations to the right of the D5 branes. Either the two
$(p,q)$ branes merge to form a $(p^\prime, q^\prime)$ brane, or they intersect.
Both of these configurations differ from  $(p,q)$-webs of fivebranes studied
in \cite{kol} by the presence of additional branes.
An analysis of these configurations is beyond the scope of this paper.
Here we simply note that no such configurations 
arise in brane constructions that give rise to asymptotically free
theories with stable vacua. Thus, the brane configuration in Fig.~\ref{fig1}
gives rise to the expected gauge theories for a total number of flavors
between $N_c$ and $3N_c$.
Repeating this analysis for a general brane configuration that satisfies
our consistency conditions yields a set of inequalities that give the
constraint $n_1+l_3 \le 2N_c$. This implies that in the general case, the
$(p,q)$ branes also do not intersect unless the total number of flavors
exceeds $3N_c$.
Thus the consistency conditions pick out those
theories that have a stable vacuum and are asymptotically free.

The consistency conditions help resolve some puzzling
aspects of the theories one can construct at zero coupling.  For instance, 
consider the theory obtained by setting $m_1=l_1=N_f>N_c$ and all other
boxes empty. This theory is forbidden by our consistency conditions, but
at zero coupling we expect to get an anomaly-free $SU(N_c)$ with $N_f$ flavors. 
For $N_f > N_c$ there should be a brane motion that yields Seiberg's duality.
In order to move the two NS branes
past each other, we connect $N_c$ D5 branes across the left NS brane so that
they extend from the right NS brane to infinity in the $-x_6$ direction.
Then we move the two NS branes to different $x_7$ positions, move them 
past each other in the $x_6$ direction and bring them back to the same
$x_7$ location. The result of this operation is a brane configuration
of the kind shown in Fig.~\ref{fig1} with $m_1=l_1=l_2=N_f$ and gauge group
$SU(N_f-N_c)$. According to the rules in \cite{hz} this theory is anomalous,
since it has $2N_f$ fundamentals and $N_f$ anti-fundamentals. Instead of the
expected Seiberg-duality \cite{ads,seiberg}, we found a brane motion that
transforms an anomaly free theory into an anomalous one. What went wrong? At
finite coupling
we see that this brane configuration is inconsistent but even if it were
consistent, the motion of the two NS branes past each other is possible 
only if the NS' branes are planes. This requires that the same number of
D5 branes end on each NS' from above and below. In other words, we must
require that $n_1=m_1=l_1$, $n_2=N_c=l_2$ and $n_3=m_3=l_3$. 
While these conditions ensure that this brane motion does not create anomalous
theories, it does not yield the expected result for the dual description of
the $SU(N_c)$ gauge theory.

To see this in some more detail, consider the theory with $n_1=m_1=l_1=N_f$
and $n_2=l_2=N_c$. In this case the two NS' branes are planes, so there is
no impediment to moving the NS branes in the $x_7$ direction. If we repeat
the brane motion described above, we get a theory with $n_1=m_1=l_1=N_f$,
$n_2=l_2=N_f-N_c$ and gauge group $SU(N_f-N_c)$. The original theory had
$N_f+N_c$ flavors while the `dual' has $2N_f-N_c$ flavors. Also, instead
of the expected gauge group $SU(N_f)$ the brane motion yields $SU(N_f-N_c)$.

\begin{figure}
\begin{center}
\centerline{\epsfxsize=7.cm  \epsfbox{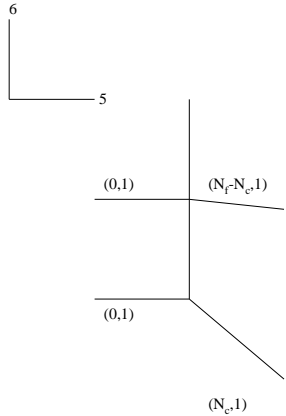}}
\caption{A section in the $x_5-x_6$ plane showing the $(p,q)$ charges
of the branes. The $(0,1)$ branes are the two NS branes and the vertical line
represents the D5 branes. The duality motion corresponds
to exchanging the $x_6$ positions of the two NS branes.  }
\label{fig3}
\end{center}
\end{figure}
\begin{figure}
\begin{center}
\centerline{\epsfxsize=7.cm  \epsfbox{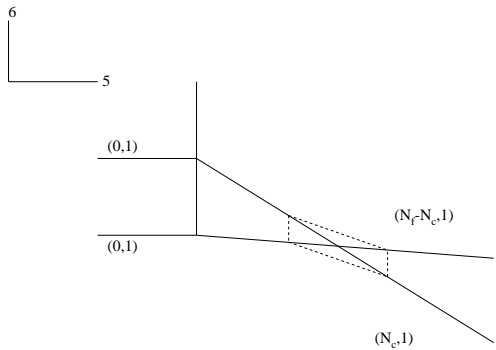}}
\caption{ The brane configuration after the duality motion. The solid lines
to the right of the D5 branes are the intersecting $(p,q)$ branes and the
dashed lines show one possible face.}
\label{fig4}
\end{center}
\end{figure}

In this analysis we did not take the bending of the NS branes into account. 
Fig.~\ref{fig3} shows a section in the $x_5-x_6$ plane of the initial brane
configuration. The brane motion described above corresponds to exchanging
the $x_6$ positions of the two NS branes. The solid lines in Fig.~\ref{fig4}
show the final configuration. In order to obtain brane 
configurations that give rise to asymptotically free field theories, we require
that the NS branes do not intersect before the brane motion.
In the initial configuration this restricts $N_f$ to the range
$0\le N_f\le 2N_c$. As shown in Fig.~\ref{fig4}, exchanging the $x_6$ positions
of the NS branes requires that the $(p,q)$ branes intersect when they
coincide in the $x_7$ direction. This is true for all values of $N_f$ except
$N_f=2N_c$, which corresponds to parallel NS branes to the left and 
parallel $(p,q)$ branes to the right of the D5 branes.

It was shown in Ref.~\cite{kol} that an intersection of $(p,q)$ branes
can be viewed as a special case of a $(p,q)$ web theory with four
external branes.
We can replace the intersection point by a finite size face, because
changing the size of a face in a $(p,q)$-web is a flat
direction corresponding to a modulus in the five-dimensional
theory on the intersection of the fivebranes.
The resulting brane configuration is shown in
Fig.~\ref{fig4}, where the dashed lines show one possible face that could
replace the intersection of the $(p,q)$ branes. We can increase the size of
the face until one edge of it touches the D5 branes. The massless states from
strings stretched between the edge and the stacks of D5 branes are degrees of
freedom in the four-dimensional field theory.
An analysis of these effects is beyond the scope of this
note, but these considerations suggest that we may need to modify the counting
rules proposed in Ref.~\cite{hz}, if there are extra intersections
of $(p,q)$ branes. One possibility is, that the configurations before
and after the brane motion give rise to the same field theory after the
states from the intersection are included.

\section{conclusions}

We analyzed the type IIB construction of four-dimensional ${\cal N}=1$ $SU(N_c)$
gauge theories at non-zero string coupling.
The condition that no further supersymmetries be broken determines 
the angles at which NS and D5 branes can intersect. Based on these conditions
we obtained consistency requirements that constrain which gauge theories
can be constructed in this approach. We found that the gauge theories are
guaranteed to be non-anomalous and that the total number of flavors is
restricted to the range $[ N_c, 3N_c]$. If the total number of flavors
exceeds $3N_c$, extra intersections between $(p,q)$ branes appear and it is
not clear what effect this has on the field theory. The IIB
construction without the extra intersections automatically gives rise only to
asymptotically free theories with stable vacua.

We also demonstrated that brane motions analogous to the IIA brane motions
that provide a string theory realization of Seiberg's duality, naively
give results that contradict the field theory expectations. At
finite string coupling, these brane motions require that new intersections
of $(p,q)$ branes appear. We argued that there may be additional massless
states in the four-dimensional field theory that arise from these intersections.

The results presented here are only a first step toward a more complete
analysis. In particular, it would be very interesting to analyze theories
with intersecting $(p,q)$ branes. This may shed some light on how to construct
Seiberg's duality in the IIB framework. An analysis of the flat directions
of these brane configurations would be useful. The zero coupling results
may get modified when the bending of branes and the possibility of replacing
intersections with faces is taken into account.  Another interesting question
is, how to modify the consistency conditions to incorporate orientifolds.
We hope to address some of these questions in the future.

\acknowledgments

It is a pleasure to thank Sergei Cherkis, Amihay Hanany, Anton Kapustin,
Iain Stewart,
and Lisa Randall for many helpful discussions. I am especially grateful to
This work was supported in part by the U.S.\ Dept.\ of Energy under Grant no.\
DE-FG03-92-ER~40701.

{\tighten

}

\end{document}